\documentclass[10pt,conference,a4paper]{IEEEtran}
\usepackage{cite}
\usepackage{amsmath,amssymb,amsfonts}
\usepackage{algorithmic}
\usepackage{array}
\usepackage{graphicx}
\usepackage{textcomp}
\usepackage{xcolor}
\def\BibTeX{{\rm B\kern-.05em{\sc i\kern-.025em b}\kern-.08em
    T\kern-.1667em\lower.7ex\hbox{E}\kern-.125emX}}

\usepackage{amsmath}
\usepackage{pgfplots}
\usepackage{pgfplotstable}
\usepackage{booktabs}

\pgfplotsset{width=10cm,compat=1.9, tick align=outside}
\pgfplotscreateplotcyclelist{my exotic colorlist}{%
  teal,every mark/.append style={fill=teal!40!black},mark=otimes*\\%
  orange,every mark/.append style={fill=orange!80!black},mark=square*\\%
  cyan!60!black,every mark/.append style={fill=cyan!80!black},mark=triangle*\\%
  red!70!white,mark=star\\%
  lime!80!black,every mark/.append style={fill=lime},mark=diamond\\%
  red,every mark/.append style={solid,fill=red!80!black},mark=*\\%
  yellow!60!black,densely dashed,every mark/.append style={solid,fill=yellow!80!black},mark=square*\\%
  black,every mark/.append style={solid,fill=gray},mark=otimes*\\%
  blue,densely dashed,mark=star,every mark/.append style=solid\\%
  red,densely dashed,every mark/.append style={solid,fill=red!80!black},mark=diamond*\\%
}

\begin{document}

\title{Segmentation of Macular Edema Datasets\\ with Small Residual 3D U-Net Architectures}

\makeatletter
\newcommand{\linebreakand}{%
  \end{@IEEEauthorhalign}
  \hfill\mbox{}\par
  \mbox{}\hfill\begin{@IEEEauthorhalign}
}
\makeatother

\author{\IEEEauthorblockN{Jonathan Frawley\IEEEauthorrefmark{1}\IEEEauthorrefmark{2}, Chris G.~Willcocks\IEEEauthorrefmark{1}, Maged Habib\IEEEauthorrefmark{3}, Caspar Geenen\IEEEauthorrefmark{3}, David H.~Steel\IEEEauthorrefmark{3}\IEEEauthorrefmark{4} and Boguslaw Obara\IEEEauthorrefmark{1}\IEEEauthorrefmark{2}}
\IEEEauthorblockA{\IEEEauthorrefmark{1}\textit{Department of Computer Science}, \textit{Durham University}, Durham, UK}
\IEEEauthorblockA{\IEEEauthorrefmark{2}\textit{Intogral Limited}, Durham, UK}
\IEEEauthorblockA{\IEEEauthorrefmark{3}\textit{Sunderland Eye Infirmary}, Sunderland, UK}
\IEEEauthorblockA{\IEEEauthorrefmark{4}\textit{Newcastle University}, Newcastle Upon Tyne, UK}
}

\maketitle

\begin{abstract}
This paper investigates the application of deep convolutional neural networks with prohibitively small datasets to the problem of macular edema segmentation.
In particular, we investigate several different heavily regularized architectures.
We find that, contrary to popular belief, neural architectures within this application setting are able to achieve close to human-level performance on unseen test images without requiring large numbers of training examples.
Annotating these 3D datasets is difficult, with multiple criteria required.
It takes an experienced clinician two days to annotate a single 3D image, whereas our trained model achieves similar performance in less than a second.
We found that an approach which uses targeted dataset augmentation, alongside architectural simplification with an emphasis on residual design, has acceptable generalization performance - despite relying on fewer than 15 training examples.
\end{abstract}


\section{Introduction}

The number of adults with diabetes worldwide has increased from 108 million to 422 million in the period 1980-2014~\cite{zhou2016worldwide}.
The number of affected adults worldwide is expected to rise to 592 million by 2035~\cite{guariguata2014global}.
About 25\% of people with diabetes have some form of diabetic retinopathy \cite{virgili2015optical}.
This is one of the leading causes of blindness for working-aged adults in the United Kingdom~\cite{arun2009long}~\cite{bunceleading2006}.
Diabetic macular edema is the accumulation of extracellular fluid in the retina secondary to inner retinal blood barrier breakdown associated with diabetes.
It results in retinal thickening in the important central retina and causes impaired vision.
It is the leading cause of decreased vision caused by diabetic retinopathy~\cite{ferris1984macular}.
Recent research from the United Kingdom suggests that, with effective screening, the number of cases that can be caught and treated early rises significantly~\cite{arun2009long}\cite{liew2014comparison}.
Automated detection of diabetic retinopathy has been shown to reduce the burden on screening services~\cite{usher2004automated}.

Optical coherence tomography (OCT) is a non-invasive, high-resolution imaging technique that uses infrared light to provide 3D imaging of the retina~\cite{Trichonasii24}.
OCT is capable of generating high-resolution, 3D images of the retina~\cite{hee1995optical}.
It is now the de facto standard tool for diagnosing multiple retinal and macular diseases, including macular edema.

Ophthalmologists currently use OCT scans to analyse the progression of macular edema both qualitatively and quantitatively.
Although quantitatively the thickness of the retina can be measured relatively easily, the extent and location of intraretinal edema relative to the remaining neuro-retinal tissue is of key importance in assessing prognosis and monitoring response to treatment.
This is a complex problem.
Quantification of the intraretinal fluid (IRF) is something that can be done manually, but it is a slow and error prone process.
Classical image techniques have failed to yield an automated solution to this problem.

Convolutional neural networks (CNN) are a deep learning based technique for solving many image-based segmentation problems.
Most CNNs today are applied in areas where a lot of data is available to train on.
In the case of medical images, there is often a data availability problem.
Data is of a much more highly sensitive nature than is typical for many domains and has to be anonymized, which is a non-trivial process~\cite{elemamanonymising2015}.
The annotation of this data with ground truth (GT) information is also a difficult, time-consuming task.
It takes an experienced clinician between 30-45 minutes to annotate each slice of the OCT image due to ambiguity, shadowing, and often there being no clear edges in intensity to follow.
With the typical scan consisting of up to one hundred individual slices at 30-120 microns separation, it can take two days for each image to be annotated.
These challenges together limit the amount of data that can be gathered.
The requirement for deep learning processes to work with small datasets is therefore of great importance in the field of medical imaging.

The U-Net CNN architecture~\cite{ronneberger2015u} represented a step forward for the accuracy of deep learning-based biological image segmentation.
It takes as input a 2D medical image and outputs a segmentation probability map.
This represents a set of probabilities $p \in [0,1]$ of each pixel being a part of the segmented region.
It comprises a series of downsampling convolutions followed by a series of upsampling mixed with 2D convolutions.
The key contribution of U-Net is the addition of \emph{skip-connections} which connect the downsampling layers with their upsampling equivalent.
This allows the model to capture fine details in the result, while the lower layers of the model will capture the general shape of the segmentation.
The combination of the two approaches has yielded very good results in a wide range of biomedical image segmentation problems~\cite{oktay2018attention}~\cite{falk2019u}~\cite{dong2017automatic}.
The 3D U-Net architecture~\cite{cciccek20163d} extends U-Net for use with 3D images by using 3D convolutions in place of 2D convolutions.
Using 3D images allows for improved segmentation as context from multiple slices aids the decision about whether an individual voxel is an object or not.

The majority of medical imaging deep learning research has involved developing segmentations for different forms of cancerous tumors and brain disease.
There has been relatively little research done on ophthalmic segmentation using deep learning~\cite{litjens2017survey}.
This is starting to change, with recent research~\cite{de2018clinically}~\cite{schlegl2018}, but it still lags behind other areas of medical imaging.

We have based one of our models on 3D U-Net with added residual blocks similar to He et al.~\cite{he2016identity}.
We also present the result of combining the above models with a Wasserstein Generative Adversarial Network (WGAN)~\cite{arjovsky2017wasserstein}, acting as a regularization approach.

We propose that using the above techniques for macular edema segmentation on a small, carefully augmented dataset yields results comparable to human performance.

\section{Method}
Segmentation involves labelling objects in an image, by assigning pixels with shared characteristics to corresponding class labels.
In our case, we wish to assign areas of IRF in an OCT image to white pixels, and non-IRF regions to black pixels.

This means we have two classes, IRF or non-IRF, which is an example of binary image segmentation:
\begin{equation}
S(x, y, z) =   \begin{cases}
    1 & S(x, y, z) \in D \\
    0 & S(x, y, z) \notin D
  \end{cases}
\end{equation}
where $D$ is the set of voxels which correspond to disease in the original image, and $x$, $y$ and $z$ represent the coordinates of that voxel~\cite{chen2017image}.

Therefore we estimate the probability of each voxel either being IRF or not, where we minimize the binary cross-entropy:
\begin{equation}
\begin{split}
    \mathcal{L}_{\text{BCE}} & = -\sum_i^{n=2} p_i \log q_i \\
    & = -(p_i \log q_i + (1-p_i) \log (1-q_i))
\end{split}
\end{equation}
where $p_i$ are the target probabilities, and $q_i$ is the output of our model.

In cases of multiple annotations per image, which we trust with equal integrity, the target probabilities $p_i$ are averaged, although we do not have any such cases except in our test set.

\subsection{Adaption of U-Net}
\label{sec:deeplearningmodel}

\begin{figure*}[ht]
  \centering
  \includegraphics[width=\linewidth]{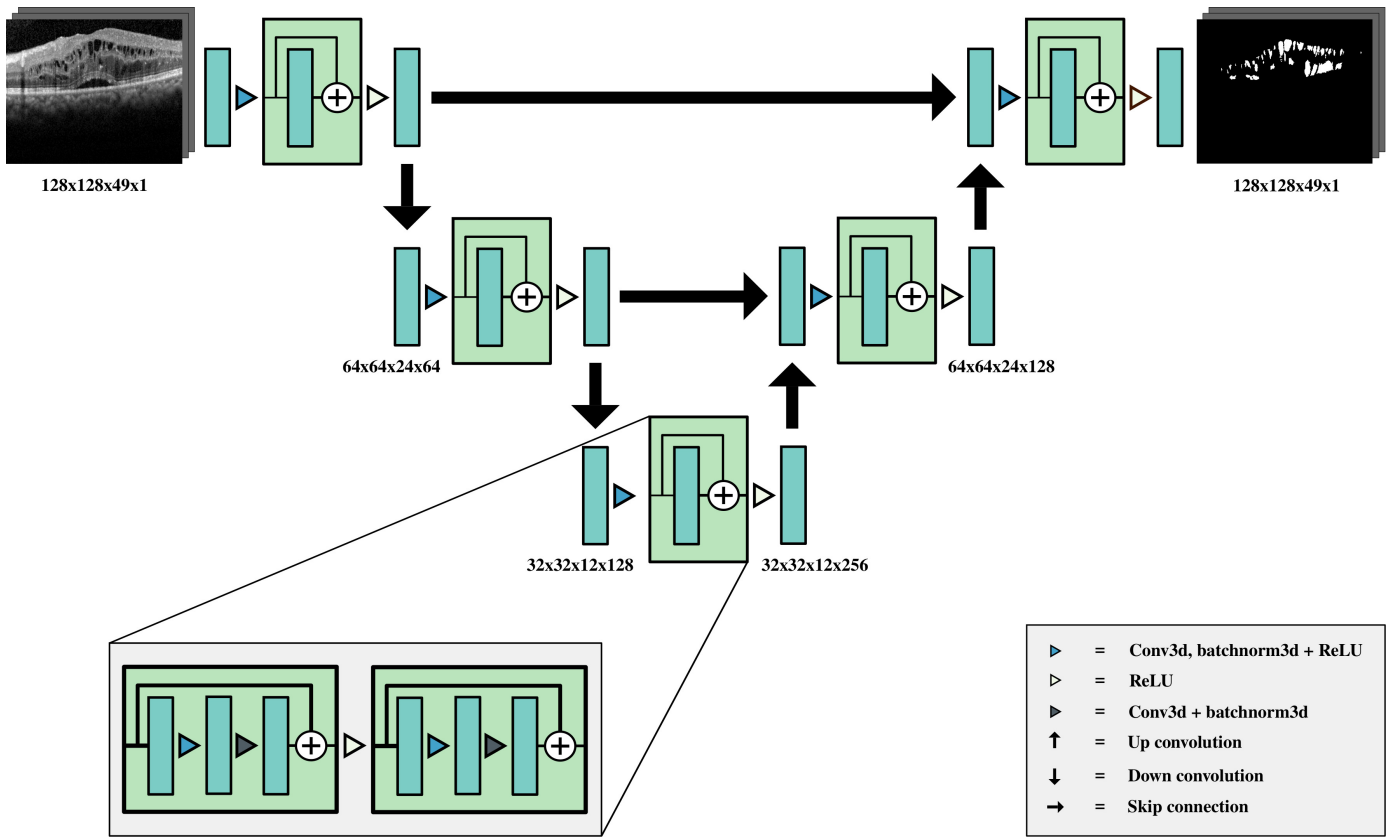}
  \caption{Small residual 3D U-Net ($M_{3}$). This model yielded the best generalization performance in our analysis. This architecture has one fewer layer than the original 3D U-Net by \c{C}i{\c{c}}ek et al~\cite{cciccek20163d} and the input and output sizes have been modified to better suit our data.}
  \label{fig:deeplearningmodel}
\end{figure*}

We investigated and designed a number of models for comparison:
\begin{enumerate}
  \setlength\itemsep{0em}
  \addtolength{\itemindent}{1em}
  \item [$M_{1}$:] Original 3D U-Net
  \item [$M_{2}$:] Small 3D U-Net
  \item [$M_{3}$:] Small residual 3D U-Net
  \item [$M_{4}$:] $M_{2}$ with WGAN
  \item [$M_{5}$:] $M_{3}$ with WGAN
\end{enumerate}

A diagram of model $M_{3}$ is shown in \figurename~\ref{fig:deeplearningmodel}.
Two residual blocks have been added to each layer.
We experimented with different model depths and found that having three layers gave the best results, while still fitting in available GPU memory.
Similarly, by experimentation, we found that the best input to all of our models is a $128 \times 128 \times 49$ image, the output is similarly a set of $128 \times 128 \times 49$ probabilities.

The WGAN models $M_{4}$ and $M_{5}$ adversarially train the U-Net against the discriminator network, such as to regularize the output to look like the same distribution as the annotations.

$M_{1}$ is a close replica of the original 3D U-Net~\cite{cciccek20163d} with batch normalization.
The input to this is a $132 \times 132 \times 116$ image and the output is a set of $44 \times 44 \times 28$ probabilities.
As the input to this model has more slices than our source images, this model is not very well suited to our dataset.

We optimize our network parameters using the Adam optimization algorithm, which is shown to give state-of-the-art performance in a number of settings~\cite{kingma2014adam}.
We considered stochastic gradient descent (SGD) as it requires less memory, but found the improvements offered by Adam to outweigh the additional memory.

\subsection{Data Augmentation in 3D}
Data augmentation is the process of expanding the training dataset by adding transformations to the inputs, artificially simulating variations that may otherwise occur naturally.
It is important that the generated data is representative of real world data.

At first, no data augmentation was performed on our dataset.
The model performed poorly on images which were at different scales to the training data.
To counteract this, we used the following transforms, all performed in 3D:
\begin{itemize}
  \setlength\itemsep{0em}
  \item Scaled up image augmentation
  \item Cropped data augmentation (equivalent to zooming in images)
  \item Elastic deformations~\cite{elasticdeformweb}
\end{itemize}
For the first two transforms, a random size is chosen to either crop or scale to.
For the scaled up case, the randomly chosen size is limited to between 1x and 4x the original dimensions of the image.
The $x{:}y$ and $x{:}z$ aspect ratios of the 3D image are preserved with scaling.
For the cropped augmentation case, a random size is chosen between 1/4 of the size and the full size of the image, again maintaining aspect ratios.
For areas outside of the boundary of the image, reflection of the original image is used.
Using a combination of these forms of augmentation, the trained model is able to cope with data at a variety of resolutions and scales.

Performing elastic deformation augmentation improves generalization performance by creating new images which are still biologically realistic~\cite{cciccek20163d}.
Parameters used for elastic deformation augmentation are $\sigma=10$ and $points=6$.
We uniformly randomly choose between these augmentation methods and performing no augmentation when creating our augmented dataset.
All of our models were trained on the result of running this augmentation on each image in our training set multiple times.

\begin{figure*}[t]
  \setlength\tabcolsep{.001\linewidth}
  \centering
  \begin{tabular}{cccccc}
    \includegraphics[width=.249\linewidth]{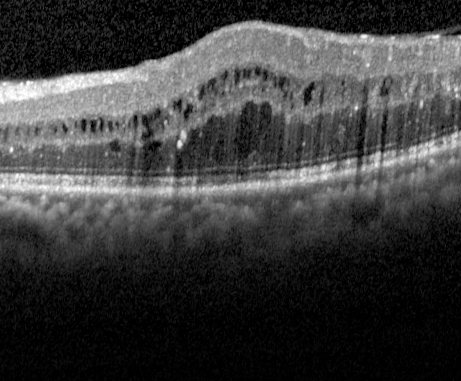} &
    \includegraphics[width=.249\linewidth]{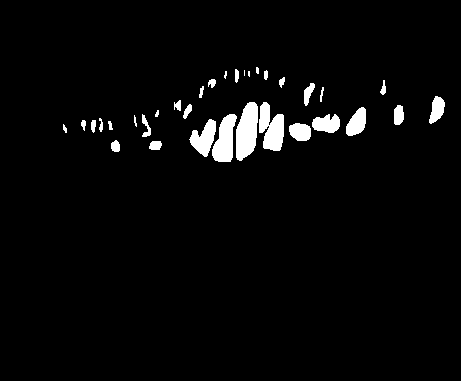} &
    \includegraphics[width=.249\linewidth]{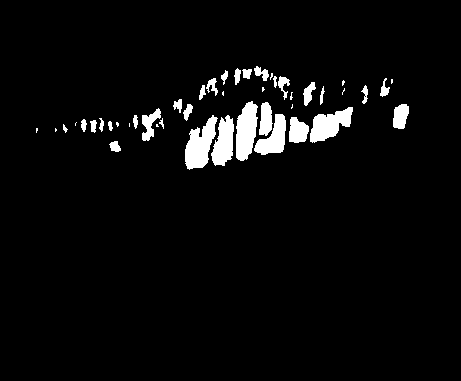} &
    \includegraphics[width=.249\linewidth]{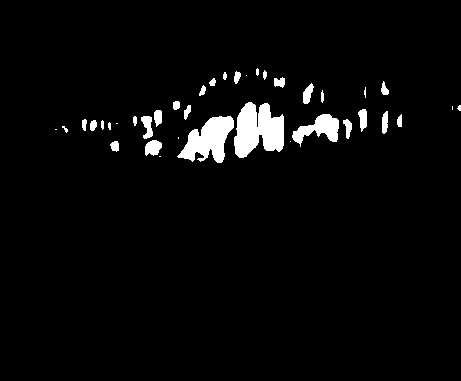} \\
    OCT scan & $A_{1}$ & $A_{2}$ & Our Prediction \\
  \end{tabular}
  \caption{A single slice of one of our unseen test OCT images alongside annotations by a clinician ($A_{1}$) and non-clinician ($A_{2}$). Comparing the manual annotations to the output of our model (on the right), it can be seen that our model captures the general structure of the IRF but fails to capture finer details from the original annotations}
  \label{fig:edemaim}
\end{figure*}

\subsection{Training}
The model described in Section~\ref{sec:deeplearningmodel} was implemented and trained using our training set.
The images were scaled down to $128 \times 128 \times 49$ in order for our model to fit in GPU memory.
In order to evaluate the model quantitatively, we scale up the image to the original size using trilinear interpolation and threshold the output probability map at $0.5$ to generate a binary image.
An example of this can be seen in the rightmost column of \figurename~\ref{fig:edemaim}.
For models $M_{2}$ to $M_{5}$, a learning rate of $1\mathrm{e}{-4}$ was used.
For regularization, we experimented with different values of weight decay for our optimizer and found that $1\mathrm{e}{-4}$ consistently resulted in the best performance on our validation set for models $M_{2}$ to $M_{4}$.
For model $M_{1}$, disabling weight decay and using a learning rate of $1\mathrm{e}{-5}$ produced the best results on our validation set.
We trained each model separately three times to assess the consistency of our results.

\section{Results}
\subsection{Qualitative Results}

\begin{figure}
    \centering
    \includegraphics[width=\linewidth]{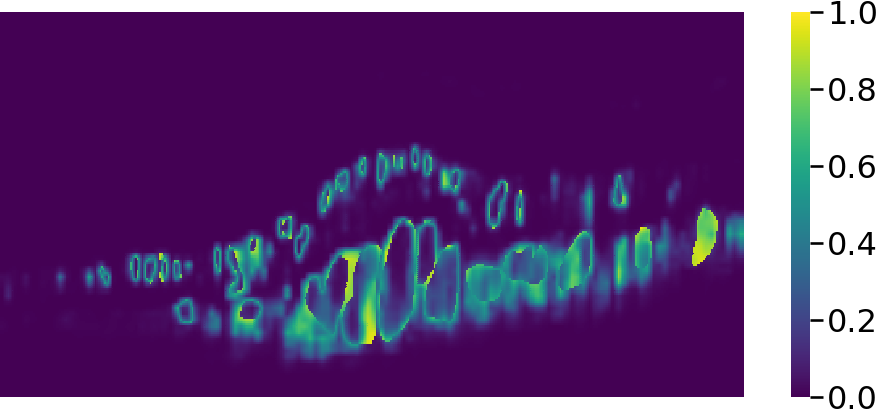}
    \caption{Heatmap of the example in \figurename~\ref{fig:edemaim}, visualising the delta between $A_{1}$'s annotation and the unthresholded model output. Brighter colours indicate areas of greater difference between the annotation and our prediction. We can see that our model tends to make most prediction errors around the edges of edema.}
    \label{fig:heatmap}
\end{figure}

The qualitative results of running the trained model are generally quite close to the ground truth, as seen in \figurename~\ref{fig:edemaim}.
\figurename~\ref{fig:heatmap} is a heatmap showing the areas of greatest difference between our model's output and the clinician's annotation.
The system tends to perform poorly in areas around the edges of edema.
This is likely due to the small size of the training set as well as the training set containing ground truths from authors with different skill levels and thresholds for delineating small areas of IRF.
The model's reduced input and output size relative to the true size of the images possibly means that it misses out on finer features in the input image.

A screenshot of a 3D viewer showing a segmented edema volume can be seen in \figurename~\ref{edema3dviewer}.
This viewer was developed to make it easier for clinicians to test and visualise the output of our model.

\subsection{Quantitative Results}

\begin{table*}[h!]
    \renewcommand{\arraystretch}{1.3}
    \centering
    \caption{Peak Jaccard Index of Tested Models against Expert Performance on the Unseen Test Dataset (Means and Standard deviations over Three Runs)}
    \setlength\tabcolsep{3.5pt}
\begin{tabular}{@{}cccccccc@{}} \toprule
& \multicolumn{5}{c}{Models} & \\ \cmidrule(r){2-6}
Author&$M_{1}$&$M_{2}$&$M_{3}$&$M_{4}$&$M_{5}$&Expert agreement\\ \midrule
$A_{1}$&0.206 (0.006)&0.545 (0.006)&\textbf{0.552 (0.011)}&0.437 (0.029)&0.441 (0.034)&0.583\\
$A_{2}$&0.225 (0.0)&0.521 (0.003)&\textbf{0.527 (0.012)}&0.452 (0.016)&0.467 (0.012)&0.583\\
  \bottomrule
\end{tabular}
    \label{modelcomparison}
\end{table*}

\begin{table*}[ht]
    \renewcommand{\arraystretch}{1.3}
    \centering
    \caption{Detailed Statistics of Best Performing Model $M_{3}$ (Means and Standard Deviations Over Three Runs)}
    \setlength\tabcolsep{3.5pt}
    \begin{tabular}{@{}cccccccc@{} } \toprule
Author&Precision&Recall&Dice similarity coefficient&Absolute volume difference&Average precision\\ \midrule
$A_{1}$&0.675 (0.013)&0.751 (0.007)&0.711 (0.01)&13783 (2099.839)&0.511 (0.013)\\
$A_{2}$&0.712 (0.006)&0.669 (0.015)&0.69 (0.01)&8731 (2099.839)&0.482 (0.014)\\
    \bottomrule
    \end{tabular}
    \label{detailedresultsbestmodel}
\end{table*}

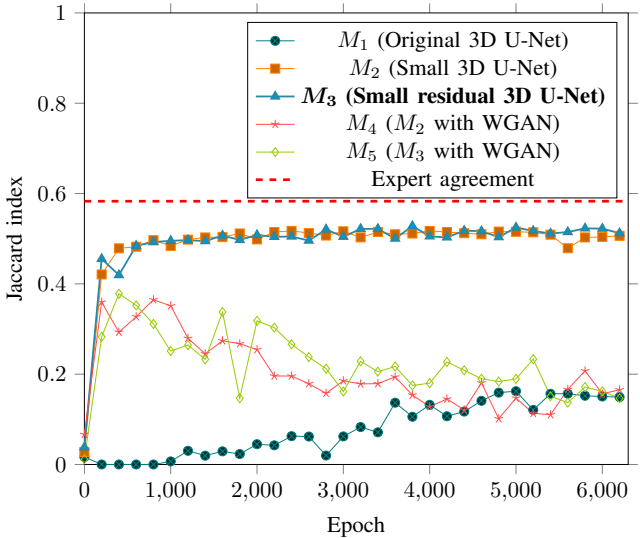
\begin{figure}[ht]
    \centering

    \begin{tikzpicture}[thick,scale=0.85, every node/.style={scale=0.8}]
    \tikzstyle{every node}=[font=\normalsize]
    \begin{axis}[
        xlabel={Epoch},
        ylabel={Jaccard index},
        xmax=6300,
        ymin=0.0,
        ymax=1.0,
        enlarge x limits=false,
        cycle list name=my exotic colorlist
        ]

    \addplot+[] table [
        x=epoch,
        y=jaccard,
        col sep=comma
        ] {./results/160/Original_U-Net_3D.csv};
    \addlegendentry{$M_{1}$ (Original 3D U-Net)}

    \addplot+[] table [
        x=epoch,
        y=jaccard,
        col sep=comma
        ] {./results/160/U-Net_3D.csv};
    \addlegendentry{$M_{2}$ (Small 3D U-Net)}

    \addplot+[thick] table [
        x=epoch,
        y=jaccard,
        col sep=comma
        ] {./results/160/U-Net_3D_Residual.csv};
    \addlegendentry{$\boldsymbol{M_{3}}$ \textbf{(Small residual 3D U-Net)}}

    \addplot+[] table [
        x=epoch,
        y=jaccard,
        col sep=comma
        ] {./results/160/U-Net_3D_GAN.csv};
    \addlegendentry{$M_{4}$ ($M_{2}$ with WGAN)}

    \addplot+[] table [
        x=epoch,
        y=jaccard,
        col sep=comma
        ] {./results/160/U-Net_3D_Residual_GAN.csv};
    \addlegendentry{$M_{5}$ ($M_{3}$ with WGAN)}
    
    \addplot[mark=none, red, dashed, very thick] coordinates {(0,0.583) (6400,0.583)};
    \addlegendentry{Expert agreement}
    
    \end{axis}
    \end{tikzpicture}

    \caption{Average Jaccard index over 3 runs on unseen test set as model is trained (higher is better). $M_{3}$ achieves the highest peak performance, reaching within 4\% of expert agreement.}
    \label{jacovertimefig}
\end{figure}

As we have only one OCT image with annotations from different authors, our comparison to human performance is limited.
For this one image, the best model achieves within 4\% of human performance as can be seen in Table~\ref{modelcomparison}.
Higher values are best for these, with $M_{3}$ achieving the best result over three runs.
$M_{2}$ has the smallest standard deviation over three runs.

Performance of the best performing model using a variety of standard image segmentation metrics is shown in Table~\ref{detailedresultsbestmodel}.
For all metrics except absolute volume difference, higher values are best.

\figurename~\ref{jacovertimefig} shows how the Jaccard index improves as the models are trained, where each epoch is 10 iterations long.
Data points are averaged every 200 epochs. $M_{3}$ reaches the highest peak while $M_{2}$ has the smoothest curve.
$M_{4}$ and $M_{5}$ both perform well below the non-GAN models, and these also take longer to train.
$M_{1}$ reaches the lowest peak performance of all models.
This could be partially explained by the significantly lower resolution output compared to the other models tested.
The residual and non-residual models' average performance is broadly similar, but the residual model has slightly better peak performance.

\section*{Implementation}
The models were implemented using PyTorch~\cite{pytorch} and were trained on $11\text{GB}-24\text{GB}$ NVIDIA Pascal and Turing architecture GPUs. Each model was trained for $6400$ epochs, which was enough for models to stop substantially improving test performance as can be seen in \figurename~\ref{jacovertimefig}.

Inference takes less than $1$s per input image using the GPU-accelerated version of our model.
PyTorch was used as it enables quick prototyping of deep learning models while also having good performance.

The Jaccard index was primarily used for evaluating the performance of our algorithm (where \emph{Pred} is our prediction and \emph{GT} is the ground truth):
\begin{equation}
  J = \frac{|Pred \cap GT|}{|Pred \cup GT|}
\end{equation}

Also known as Intersection over Union (IoU), this is a commonly used metric for comparing the similarity of two sets.
In this case, the Jaccard index represents the intersection of the model's prediction and the ground truth divided by the union of the model's prediction and the ground truth.
It was used as a key metric when evaluating the performance of our prediction, thresholded at 0.5, as it is a robust indicator of how close the resultant segmentation is to the ground truth.

The open source software, Scikit-learn, was used to compute all metrics~\cite{scikitlearn}.

\section*{Materials}
\begin{table}
  \renewcommand{\arraystretch}{1.3}
  \centering
  \caption{Number of Annotations Per Dataset and Author}
  \setlength\tabcolsep{3.5pt}
  \begin{tabular}{@{}lccc@{} } \toprule
& \multicolumn{3}{c}{Author} \\ \cmidrule(r){2-4}
Dataset & $A_{1}$ & $A_{2}$ & $A_{3}$ \\ \midrule
Train&8&0&6 \\
Validation&0&0&2 \\
Test&1&1&0 \\
  \bottomrule
  \end{tabular}
  \label{dataset}
\end{table}

OCT images were exported from a Heidelberg SPECTRALIS HRA+OCT machine with software version 1.10.4.0.
These images were cropped to remove unnecessary information and the fundus image.
Annotations were created manually using a 3D image annotation tool, slice by slice in the $z$-dimension, by highlighting the pixels on the OCT scan which are of IRF.
Table~\ref{dataset} shows how (image, annotation) pairs were divided into training, validation and test sets based on who authored each ground truth.
Due to the many hours it takes to annotate a single image, our dataset sizes are small.
Also note the imbalance of annotations for each author.
$A_{1}$ is a clinician, $A_{2}$ and $A_{3}$ are non-clinician image and data experts. 
We use the image with multiple annotations as our test set in order to be able to compare our model against expert agreement.
All images and ground truths at full size have dimensions $\text{width}=461,\text{height}=381,\text{slices}=49$.

\section{Conclusions and Future Work}
\begin{figure*}
    \centering
    \includegraphics[width=\linewidth]{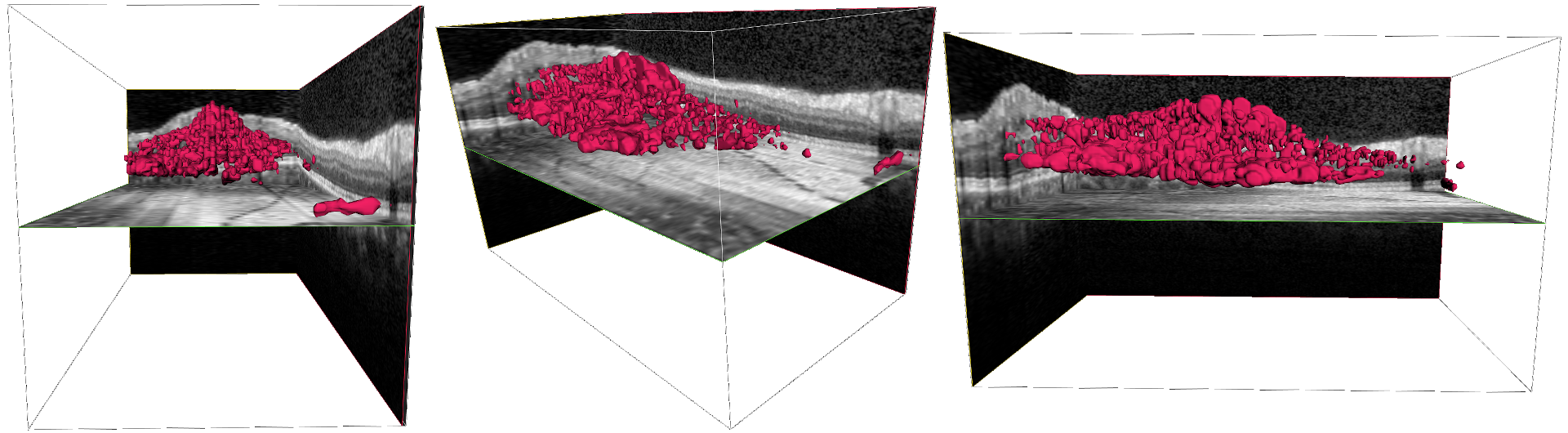}
    \caption{Model output viewed through interactive 3D WebGL-based viewer (Online platform by Intogral Ltd.).}
    \label{edema3dviewer}
\end{figure*}

It is hypothesised that our model performs so well due to its simplicity.
As we are capable of learning a working solution with fewer layers than the original 3D U-Net, we think that macular edema segmentation is well suited to a small residual U-Net architecture.

It takes days to create a single annotated 3D OCT image by hand.
This makes it infeasible for clinicians to manually create these annotations for every patient.
Our solution can generate a similar quality result automatically in less than a second.
Future work should be able to use the resultant segmentation to calculate metrics such as volume and surface area in an automated manner.
Future clinical research will be able to assess correlation of these metrics with disease progression and treatment outcomes.

As it takes so long to create annotations, increasing the size of the dataset is difficult.
We do think, however, that if more data from expert clinicians were trained on, results would continue to improve.

The model primarily makes mistakes around the edges of edema.
We believe this is partially due to the fact that the images are scaled down prior to being input to the model.
It would be useful for the model to work on the full resolution image.
This could be done using techniques similar to those used to create super-resolution images as described by Dong et al~\cite{7115171}, or by using GPUs with larger amounts of memory available.
It would also be useful to develop interactive tools that allow for quick refinement of any segmentations generated by our network.

The imaging device used also provides fundus image output.
An interesting extension of this project would be to use the fundus output along with the OCT image and see if that improves the prediction.
The fundus image could help with prediction by utilising features for maculopathy such as exudates and red lesions (microaneurysms).
The role of vessel width and geometry analysis in maculopathy prediction could be added to the OCT biomarkers to improve prediction accuracy for disease progression and response to treatment.
Recent work on deep learning models for diagnosing age-related macular degeneration (AMD) has shown that combining OCT and fundus image output can yield improved results\cite{yoo2019possibility}.

The use of another medical imaging technique known as OCT angiography (OCT-A) has shown promise in helping to diagnose diabetic retinopathy and macular edema\cite{kim2016quantifying}~\cite{mao2017optical}.
Recent work has suggested combining OCT-A, OCT and fundus images to improve the accuracy of models to diagnose AMD~\cite{vaghefi2020multimodal}.
Applying such an approach to the automated diagnosis of macular edema may help to improve prediction accuracy.

We have trained and tested images from a single device type (Heidelberg SPECTRALIS HRA+OCT).
In order to create a real-world diagnostic solution, it would be required to train and test our model on images from a variety of OCT device manufacturers.
Recent work by De Fauw et al~\cite{de2018clinically} has shown that it is possible to train a model on a single OCT device and refine it to work for another OCT device.

\section*{Conflict of Interest \& Attribution}
In accordance with his ethical obligation as a researcher, Jonathan Frawley reports that he receives funding for his PhD from Intogral Ltd.
Some of the work described was developed as part of his work as an employee at Intogral Ltd.
Data and annotations by the clinician for this project were kindly provided by Maged Habib, Caspar Geenen and David H.~Steel of the Sunderland Eye Infirmary, South Tyneside and Sunderland NHS Foundation Trust, UK.
Intogral Ltd also provided annotations, as well as providing the use of the platform for rendering objects in 3D shown in \figurename~\ref{edema3dviewer}.

\bibliographystyle{IEEEtran}
\bibliography{main}

\end{document}